\documentclass{article}

\usepackage{arxiv}

\usepackage[utf8]{inputenc} 
\usepackage[T1]{fontenc}    
\usepackage{hyperref}       
\usepackage{url}            
\usepackage{booktabs}       
\usepackage{amsfonts}       
\usepackage{nicefrac}       
\usepackage{microtype}      
\usepackage{cleveref}       
\usepackage{lipsum}         
\usepackage{graphicx}
\usepackage{natbib}
\usepackage{doi}

\usepackage{float}          
\usepackage{pbox}           
\usepackage{caption}        

\title{Brain states analysis of EEG predicts multiple sclerosis and mirrors disease duration and burden}

\date{November 5, 2025}

\newif\ifuniqueAffiliation

\ifuniqueAffiliation 
\author{ \href{https://orcid.org/0000-0000-0000-0000}{\includegraphics[scale=0.06]{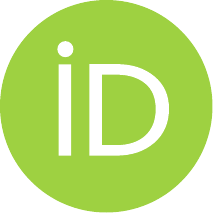}\hspace{1mm}David S.~Hippocampus}\thanks{Use footnote for providing further
		information about author (webpage, alternative
		address)---\emph{not} for acknowledging funding agencies.} \\
	Department of Computer Science\\
	Cranberry-Lemon University\\
	Pittsburgh, PA 15213 \\
	\texttt{hippo@cs.cranberry-lemon.edu} \\
	\And
	\href{https://orcid.org/0000-0000-0000-0000}{\includegraphics[scale=0.06]{orcid.pdf}\hspace{1mm}Elias D.~Striatum} \\
	Department of Electrical Engineering\\
	Mount-Sheikh University\\
	Santa Narimana, Levand \\
	\texttt{stariate@ee.mount-sheikh.edu} \\
}
\else
\usepackage{authblk}

\newbox{\orcid}\sbox{\orcid}{\includegraphics[scale=0.06]{orcid.pdf}} 
\author[1,6]{\href{https://orcid.org/0000-0002-0187-3183}{\usebox{\orcid}\hspace{1mm}István Mórocz\thanks{\texttt{pistikem@gmail.com}}}}%
\author[2]{\href{https://orcid.org/0000-0002-5665-7285}{\usebox{\orcid}\hspace{1mm}Mojtaba Jouzizadeh\thanks{\texttt{mojtaba.jouzizadeh@uottawa.ca}}}}%
\author[3]{Amir Hossein Ghaderi\thanks{\texttt{amirhossein.ghaderi@ucalgary.ca}}}%
\author[4]{Hamed Cheraghmakani\thanks{\texttt{hcheraghmakani@gmail.com}}}%
\author[4]{Seyed Mohammad Baghbanian\thanks{\texttt{mohammadbaghbanian@gmail.com}}}%
\author[5]{Reza Khanbabaie\thanks{\texttt{rezakhan@uottawa.ca}}}%
\author[6]{\href{https://orcid.org/0009-0004-2440-9422}{\usebox{\orcid}\hspace{1mm}Andrei Mogoutov\thanks{\texttt{andrei.mogoutov@noisis.com}}}}%

\affil[1]{McGill University, Montreal Neurological Institute, IPMSA International Progressive MS Alliance, Montreal, QC, Canada}
\affil[2]{University of Ottawa, Department of Neuroscience, Ottawa, ONT, Canada}
\affil[3]{Department of Psychology and Hotchkiss Brain Institute, University of Calgary, Calgary, Canada}
\affil[4]{Department of Neurology, Faculty of Medicine, Mazandaran University of Medical Sciences, Sari, Iran}
\affil[5]{Department of Physics, University of Ottawa, Ottawa, ON, Canada}
\affil[6]{Noisis Inc., Montreal, QC, Canada}
\fi

\renewcommand{\shorttitle}{EEG Brain States Analysis in MS}

\hypersetup{
pdftitle={A template for the arxiv style},
pdfsubject={q-bio.NC, q-bio.QM},
pdfauthor={David S.~Hippocampus, Elias D.~Striatum},
pdfkeywords={multiple sclerosis, EEG, brain states analysis, machine
  learning, monitoring, disease burden},
}

\usepackage{datetime2} 
\hypersetup{
  colorlinks   = true, 
  urlcolor     = blue, 
  linkcolor    = blue, 
  citecolor    = blue  
}

\begin{document}

\maketitle

\begin{abstract}

\textbf{Background:} Any treatment of multiple sclerosis should
preserve mental function, considering how cognitive deterioration
interferes with quality of life.  However, mental assessment is still
realized with neuro-psychological tests without monitoring cognition
on neurobiological grounds whereas the ongoing neural activity is
readily observable and readable.

\textbf{Objective:} The proposed method deciphers electrical brain
states which as multi-dimensional cognetoms quantitatively
discriminate normal from pathological patterns in an EEG.

\textbf{Method:} Baseline recordings from a prior EEG study of 88
subjects, 36 with MS, were analyzed.  Spectral bands served to compute
cognetoms and categorize subsequent feature combination sets.

\textbf{Result:} The brain states predictor correlates with disease
burden and duration.  Using cognetoms and spectral bands, a
cross-sectional comparison separated patients from controls with a
precision of 85\% while using bands alone arrived at 79\%.

\textbf{Conclusion:} We demonstrate the efficiency of the quantitative
data-driven method based on brain states analysis by contrasting EEG
data of patients with MS and healthy subjects.  The congruity with
disease severity and duration is a neurophysiological indicator for
disease accumulation over time.  We discuss potential applications of
the approach for the monitoring of disease time course and treatment
efficacy in longitudinal clinical studies in psychiatry and neurology.

\end{abstract}


\section{Introduction}

Mental illness is an enormous burden for societies world\-wide
\citep{Smith2011,Vigo2016}.  Above all it is the decay of cognitive
functioning, the essence of human reasoning, that turns a mental
illness into a devastating and tangible experience for patient and
surrounding alike. Nonetheless diagnosis and staging of diseases still
rely on customary clinical, laboratory and imaging tests by which
cognitive integrity is not assessed on biological grounds.  Electrical
scalp recordings have existed for a century and have been used decades
ago as diagnostic evidence in multiple sclerosis, although the
analytic methods lacked the quantitative aspect needed to relate brain
measures to disease time course or treatment outcome.  We believe that
a modern data-driven functional brain mapping analysis approach using
non-invasive EEG data will meaningfully enhance the prognostic power
for clinical trials and facilitate the routine assessment of drug
efficacy and safety through objective measures for a changing
cognito-electrical substrate.

The central nervous system (CNS) is an electrical organ and invokes
with each mental act a plethora of neural activities instantly
reflected in the electrical brain space and recordable on the skull
surface with electrode arrays \citep{Buzsaki2005c}.  Electrical states
of brain functioning, coined \emph{cognetoms}, change at any given
moment \citep{Koenig1999,Samdin2017} and are traced with physiological
measures while anatomical brain lesions evolve slower over days or
weeks and are monitored with MRI exams
\citep{Eshaghi2021,Tommasin2021,Lazzarotto2024,Goebl2025}. The crucial
difference is in the immediacy and temporal accuracy of electrical
measures and the dispersed soft demarcation of electrical phenomena on
the skull surface that mirror underlying distributed brain activity.
These diffuse and perpetually evolving cloud-like arrangements are due
to oscillating neural networks and reflect equally housekeeping and
cognitive neural processes \citep{Buzsaki2005c}.

MS is a chronic, often relapsing‑remitting, inflammatory disease with
multiple foci of demyelination disseminated over space and time
throughout the CNS.  It can lead to impaired cognitive reserves and
capacity \citep{Maggi2025,Pauletti2025} disability
\citep{CoboCalvo2025} and early death.  Northern countries show a high
frequency for MS, with the highest prevalence in Canada of 0.3\%
\citep{Amankwah2017}.

Recent advances in pharmaceutical research for aggressive
disease-mo\-di\-fy\-ing treatments lead to a surge of novel drugs for
MS over the last two decades.  Currently, there are 18
disease-modifying therapies approved by Health Canada (relapsing
remitting form), while heaps of new compounds are being tested in
clinical trials for clinical and MRI outcomes.  The desired effect of
drug therapy on cognitive preservation \citep{Falet2022}, however, is
still conventionally assessed with questionnaires and
neuro-psychological tests, although a recent upswing of publications
using functional brain measures in MS is observable
\citep{Kisler2020,Paolicelli2021,Santinelli2021,SarriasArrabal2021,Mohseni2022,Sorrentino2022}.

For this work we propose an extension for the concept of brain states
analysis \citep{Janoos2011,Morocz2012,Janoos2013}.  Instead of the
statistical principles based on hidden Markov model (HMM) and
functional MRI data we developed a novel method for brain states
detection and analysis of EEG signals taken from a prior
\emph{cross-sectional} EEG study about patients with multiple
sclerosis and healthy subjects \citep{Jouzizadeh2021}.

\begin{figure}
  \setlength{\fboxsep}{0pt}%
  \setlength{\fboxrule}{0pt}%
  \centering
  \includegraphics[scale=.8]{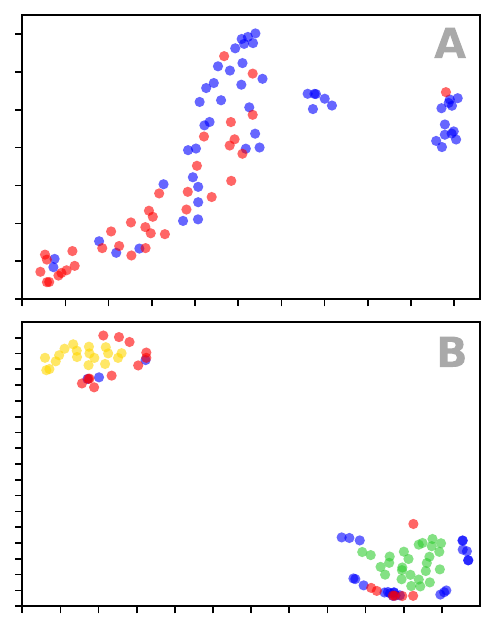}
  \caption{\textbf{(A)} Result of dimension reduction with
    \emph{un-supervised} version of UMAP manifold learning algorithm.
    The sequence of brain states for each individual data is embedded
    to a two-dimensional space.  Each point corresponds to one
    individual, red color with MS patients, blue with control
    subjects.  \textbf{(B)} Result of dimension reduction with
    \emph{supervised} version of UMAP manifold learning algorithm.
    The sequence of brain states for each individual data is embedded
    to a two-dimensional space.  Each point corresponds to one
    individual.  Green corresponds with the train subset of control
    subjects and blue with the test subset, then yellow with train
    subset of MS patients and red with the test subset.  In both
    plots, there is no meaning to axis orientation and values, only
    the relative position of points in two-dimensional space is
    relevant as well as the density distribution through clustering.}
  \label{F1}
\end{figure}

\section{Materials \& Methods}

\subsection{Subjects and EEG recording} 

The EEG data of 88 subjects were taken from an earlier project.
Participant criteria and recording sessions have been described in
detail in the original publication \citep{Jouzizadeh2021}.
Participants signed consent forms for all procedures and data analyses
in accordance with the Declaration of Helsinki.  The anonymized data
of 36 subjects diagnosed with MS (23 female, 13 male, mean age (29.6
years, age range 20-53 years) and 52 healthy volunteers (25 female, 27
male, mean age 20.7 years, age range 18-31 years) were taken into
consideration for this data analysis project (see tables in
{\textsf{\textbf{\ref{ST2}}}} and {\textsf{\textbf{\ref{ST3}}}} for
subject information).  The mean of the Expanded Disability Status
Scale (EDSS) scores for MS patients at EEG acquisition time was 1.96
(EDSS score range 0-7).  The diagnosis of MS was based on the revised
McDonald criteria for relapsing-remitting MS (RRMS) as used at
Boualicina Hospital, Sari, Iran.  The mean disease duration after
confirmed diagnosis was at EEG recording time 5.1 years (range 0-18
years).  Resting-state EEG was recorded for 5 minutes for each
condition, eyes-open (EO) and eyes-closed (EC), using a Brainmaster
Discovery24 data acquisition system and 19 scalp electrode positions
at 256\,Hz sampling rate.

\subsection{Data preparation} 

Raw binary sensor data were extracted as time stream vectors from
original EEG data files without further metadata or personal
information.  Electrode labels and anatomical positional information
played no role during data analysis.  Attention was paid to maximally
preserve the intrinsic physiological information present in the data
\citep{Delorme2023} for the brain states generation phase further
downstream in the processing pipeline.  EEG data were taken as is
without applying common manual or automated data filtering, cleansing
and artifact removal techniques for noise reduction, faulty
electrodes, eye blink, cardiac or muscle activity, and hence without
introducing data pre-processing related artifacts.

\subsection{Brain states detection (BSD)} 

We were inspired by the similarity of the brain states analysys method
we had developed for functional MRI data \citep{Janoos2011,Morocz2012}
and the concept of microstates analysis for EEG data introduced by
Lehmann et al \citep{Lehmann1987}.  Both approaches aim to transform
the complex original data into a sequence of discrete, quasi-stable
meaningful patterns, also called states.  It was shown that the
distribution of these states may play a biomarker role in various
psychiatric \citep{Koenig1999,Janoos2013} and neurological conditions
including MS \citep{Gschwind2016}.  In the current work, however, a
brain state is a typical configuration of spatial activity without
actual knowledge of spatial coordinates of electrodes.

We earlier defined brain states, or \emph{cognetoms}, as snapshots of
repeatable patterns of highly correlated signals also understood as
moments of distributed brain activity of a given time point
\citep{Janoos2011,Morocz2012}.  In other words, states are defined as
patterns of statistical dependencies or spatio-temporal correlations
between brain regions measured across EEG electrodes.  The states
serve as lowest-level elements, or level-0 cognetoms, quasi as
alphabet \emph{letters} or building blocks to use in future
aggregation cascade models of ever increasing cognetom complexity and
abstraction level.

In a \emph{first} move a limited set of eight brain states are
computed.  Raw EEG signals from each channel are transformed into
sequences of overlapping spectrograms using the Fast Fourier Transform
(FFT) implementation from the cusignal.spectrogram library
\citep{cuSignal2021}.  The recordings are segmented into data epochs
where each individual window, a brain state, consists of 500 EEG
frames (scan frequency at 256Hz, epoch length 1.953sec), and are
described by five EEG frequency band amplitudes (delta: 0-4Hz, theta:
4-8Hz, alpha: 8-14Hz, beta : 14-32Hz, gamma: 32-45Hz).  While window
size and number of states are varied as technical optimization against
detection performance, its length of 2s approximates the scan
acquisition times needed for a single multi-slice T2* weighted brain
shot typical for functional MRI studies \citep{Morocz2012,Janoos2013}.
To reduce dimensionality and capture coherent spatial patterns, we
grouped EEG channels based on the similarity of their time series.
Pairwise distances between channels were computed, followed by
hierarchical clustering to identify groups of similar channels.  For
the feature vector construction, we computed at each time step the
mean signal intensity across predefined frequency bands within each
cluster of grouped channels.  This resulted in a concatenated vector
representation summarizing the spectral profile of the EEG activity at
each moment.  For the identification of brain states, we applied
k-means clustering to the resulting vectors to identify a set of
prototypical channel-spectral configurations, which we interpret as
data-driven representations of transient brain states.

In a \emph{second} move, for the temporal coding of brain states, or
brain states sequence, the best match distribution of brain states is
detected for each time point along the EEG data time axis with weights
shown for each brain state.  Each time step in the original sequence
is assigned to its nearest brain state cluster, producing a simplified
temporal sequence of discrete brain state codes, each associated with
signal intensities across frequency bands.  In other words, our
approach translates the multi-channel raw EEG signal of each subject
into a \emph{sequence} of brain states as \emph{qualitative} feature
with the associated weights as numerical \emph{quantitative} feature.

\subsection{Brain states analysis (BSA)} 

In order to capture differences between the EEG data of healthy
persons and patients with MS two approaches were applied : manifold
learning and catboost classifier.  For the subject-level feature
extraction, 80 distinct feature combinations were built and are
described as combinations of 8 brain state weights, two experimental
recording conditions (eyes-open and eyes-closed) and five EEG
frequency band amplitudes as mentioned.  This compressed
representation enabled further statistical analysis.  For each
subject, we computed distributions over brain states and average
signal intensities per state.  We used a fixed-length feature vector
for each subject, defined by the combination of brain states (8),
frequency bands (5), and experimental conditions (2), for the
downstream comparative analyses.

The \emph{first} approach, manifold learning, reduces the
dimensionality of the individual data \citep{McInnes2018}.  The
previously prepared sequence of brain states and weights are used to
embed the individual data into a two-dimensional space.  Each point in
that space corresponds with one person.  First we applied the
\emph{non-supervised} version of the UMAP algorithm for which the
results reveal a distinct trend separating control from MS cases
(upper figure {\textsf{\textbf{\ref{F1}}}}).  Next we applied the
\emph{supervised} version by splitting the data into train (60\%) and
test (40\%) subsets.  The trained model produced the two-dimensional
embedding with high separability between the two subject cohorts
(lower figure {\textsf{\textbf{\ref{F1}}}}).

Our \emph{second} approach used the catboost classifier model
\citep{Dorogush2018} which applied a 30-fold cross-validation to split
the data into train (60\%) and test (40\%) subsets, then trains the
model, and last evaluate the quality of the model on the test data
subset.  Only the binary labels for MS and non-MS were used for model
training.  All other clinical or demographic information was taken
into account only for the post-analysis procedures.  The quality of
the model is described by precision, recall, f1-score and support (see
table {\textsf{\textbf{\ref{T1}}}}).

\section{Results}

\subsection{Benefit of the brain states approach} 

Our very first implementation of the brain states approach for EEG
analysis shows a successful performance for detecting MS in common
clinical low-resolution EEG data sets at a high precision of 85\%.  We
demonstrate a direct benefit for including brain states features in
the analysis as compared to taking into account only the frequency
band features typically used for EEG analysis (table
{\textsf{\textbf{\ref{T1}}}}).  A benchmark for the two EEG analysis
approaches show a performance boost of 6\% for distinguishing EEG data
sets of patients with MS versus healthy subjects in favor of the
combined method.  The precision for detecting MS using frequency bands
alone reaches 79\% while adding information about brain states arrives
at 85\%.


\begin{table}[!ht]
  \vspace{5mm}
  \small\sf\centering
  \begin{tabular}{lllll}
    \toprule
    separation based on                 & precision     & recall        & f1-score      & support \\
    \midrule
    frequency bands alone               & 0.79          & 0.78          & 0.78          & 36 \\
    frequency bands \& brain states     & 0.85          & 0.84          & 0.84          & 36 \\
    \bottomrule
  \end{tabular}\\[10pt]
  \caption{Performance of MS detection with catboost classifier.  We
    compare the classifier performance for two scenarios : first the
    features are defined as frequency bands and experimental
    conditions such as eyes open and eyes closed.  In the second
    scenario the features are defined as a combination of brain
    states, frequency bands and experimental conditions.  We applied a
    30-fold cross-validation to split the data into train (60\%) and
    test (40\%) subsets of the data.  Taking into account the brain
    states augments the precision, recall and f1-score for
    distinguishing EEG data sets in patients with MS \emph{vs} healthy
    subjects - as compared to taking into account the frequency band
    features alone.}
  \label{T1}
\end{table}

\subsection{Specific feature combinations for MS detection} 

\begin{figure}[!ht]      
  \setlength{\fboxsep}{0pt}%
  \setlength{\fboxrule}{0pt}%
  \centering
  \includegraphics[scale=.6]{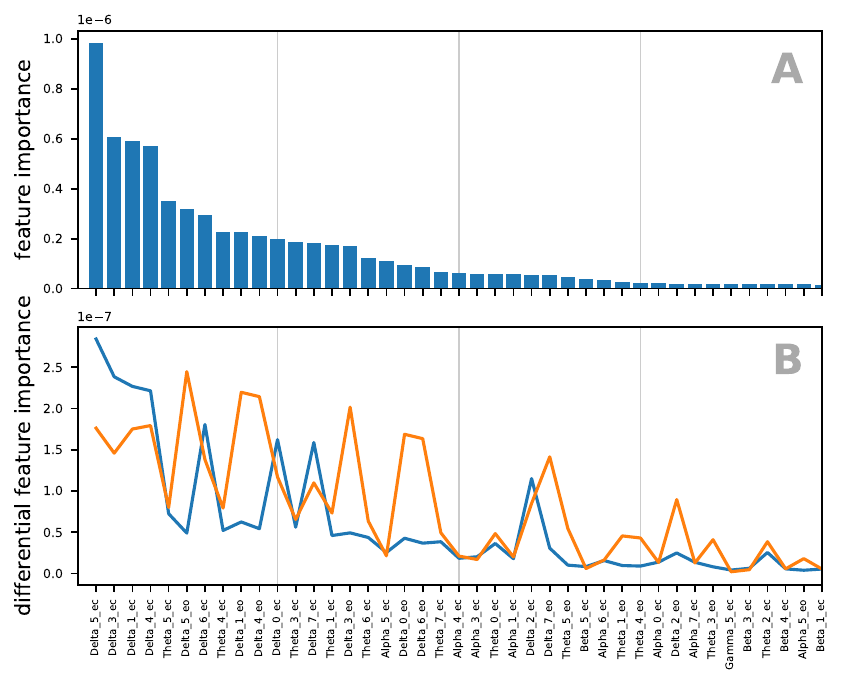}
  \caption{Specific feature analysis for MS detection.  The relative
    importance of feature combination (brain states \emph{vs}
    frequency bands \emph{vs} experimental conditions) is shown.  For
    example the first feature combination is called `Delta\_5\_ec'
    which stands for `Delta' frequency band, brain states no.\ `5',
    and `ec' for the eyes closed condition.  \textbf{(A)} The upper
    figure part depicts ordered feature combinations (x-axis)
    \emph{vs} feature importance (y-axis).  \textbf{(B)} The lower
    part shows the comparison between the control \emph{vs} the
    patient cohort regarding feature importance.  The blue line stands
    for the control group, the orange line for MS.  Only the 40 top
    most important feature combinations are shown.}
  \label{F2}
\end{figure}

The relative importance of 80 feature combinations (8 brain states x 5
frequency bands x 2 experimental conditions) that separate the two
cohorts into healthy vs MS is outlined in figure
{\textsf{\textbf{\ref{F2}}}}.  To reduce the risk of overfitting only
the first 15 feature combinations of high importance including both
eyes-open and eyes-closed conditions and the delta frequency band were
selected that appear to drive the separation between healthy and MS
subjects.  Averaged frequency spectra for both cohorts are shown as
graphs in {\textsf{\textbf{\ref{SF4}}}}.

\subsection{Disease predictor compared to clinical data} 

\begin{figure}[!ht]
  \setlength{\fboxsep}{0pt}%
  \setlength{\fboxrule}{0pt}%
  \centering
  \includegraphics[scale=.77]{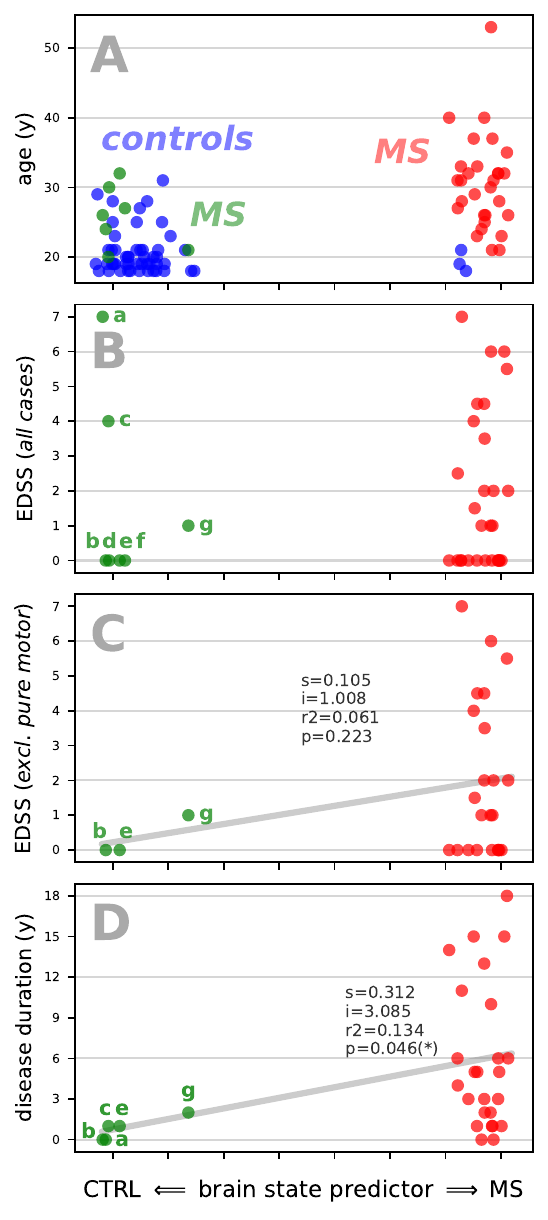}
  \caption{Distinction of control subjects from MS patients through
    brain states prediction.  Disease predictor values correspond to
    the rotated coordinates as seen in
    {\textsf{\textbf{Fig.\ref{F1}B}}} using principal components of
    the point distribution with the predictor value considered as
    primary principal component.  \textbf{(A)} The \emph{top} graph
    illustrates age distribution as a function of the MS predictor
    scale.  It indicates no observable effect of age on MS prediction.
    The control outliers on the right side (\emph{false positives in
    blue}) do not have higher ages than the controls on the left side
    (\emph{true negatives}).  Conversely, the MS cases on the left
    side (\emph{false negatives in green}) do not have a lower age
    distribution than the \emph{true positives} on the right side.
    \textbf{(B)} The \emph{second} graph illustrates this separability
    in the MS cohort plotted against EDSS measures.  Patients are
    shown as red or green dots as in the top graph.  The seven
    outliers on the left are marked with green characters (a-g) for
    further discussion under Results.  \textbf{(C)} The \emph{third}
    graph plots separability against EDSS measures only for MS
    patients having symptoms other then motor symptoms.  For these 26
    patients a linear regression is shown as gray line.  \textbf{(D)}
    The \emph{bottom} graph plots separability against disease
    duration in years.  Six patients (including d,f) are excluded for
    no data on disease duration.  Linear regression is shown as gray
    trend line.  The source for the post-analysis correlation
    calculation is in the values for the brain state predictor space
    and EDSS or disease duration.  The open-source python module
    linregress from scipy served for calculation
    \citep{Virtanen2020}.}
  \label{F3}
\end{figure}

The brain states analysis technique separates with surprising ease and
high probability healthy controls from patients with MS (figures
{\textsf{\textbf{\ref{F1}}}} and {\textsf{\textbf{\ref{F3}}}}).
Considering the data to originate from single clinical-grade EEG
recordings, we therefore examined additional objective neurological
measures such as age, EDSS values or plain disease duration which we
hypothesized will correlate with our EEG analysis.  Age itself does
not indicate an observable effect on MS prediction in both cohorts
(\emph{top} graph A in figure {\textsf{\textbf{\ref{F3}}}}).  However
the next two clinical measures arguably reflect disease related
accumulation of neurological deficits.  The next three plots in figure
{\textsf{\textbf{\ref{F3}}}} imply a correlation of these two clinical
measures with the brain states predictor space, which, we surmise,
distinguishes the better the two cohorts, the more disease burden over
time accumulates in the CNS.  Patients with symptoms other than pure
motor symptoms show a subtle trend with EDSS computed on 26 such cases
(third graph C).  Regarding disease duration, however, all patients to
the far left in the `control' brain states predictor space (marked
green with letters from \emph{a} to \emph{g}) feature a short disease
duration of less than two years (mean 0.8 years) and cited motor and
sensory deficits which may originate from lesions in the spinal cord
or lower brain rather than upper brain parts.  In contrast, patients
to the right marked red, with a mean disease duration of 6.0 years,
reported also vision, balance or brain stem disturbances beside motor
and sensory issues in the limbs.  Still, those patients in red to the
right with a brief disease duration of 3 years or less, may in fact,
we hypothesize, suffer from a longer, hidden disease duration, albeit
clinically compensated, to amass enough supratentorial pathology that
finally deranges their EEG patterns.  We assume that an increase in
lesion load in the cerebrum, the principal generator for the EEG
signal, relates just vaguely and non-linearly to the emerging complex
clinical neurological deficits, be that in cognition, EDSS, or then in
form of clinical disease duration, or `true' disease onset, an unknown
entity for that matter.

\section{Discussion}

We successfully transformed a brain states hypothesis
\citep{Janoos2011} into a new analytic method for processing EEG data
using a quantitative embedding and modeling approach for brain states
and applied the new technique to separate the two subject cohorts of
an earlier EEG study about MS \citep{Jouzizadeh2021}.  We observed
several beneficial aspects when brain states analysis is added in a
context as described here.

\subsection{Implications} 

\emph{First} we demonstrate with an initial implementation of our
novel algorithm, a two-phase processing pipeline with brain states
detection and brain states analysis, that already regular clinical EEG
data clearly carry the necessary information to discriminate two
cohorts as shown here.  This is to our knowledge a first successful
demonstration of such a hypothesis-free data-driven full-brain
raw-data EEG analysis approach for MS.

\emph{Second} we state that the inclusion of brain states in the
analysis is sensible and enhances the statistical robustness of the
approach.

\emph{Third} we believe this brain states approach is easily
generalizable to other disease conditions and importantly also for the
intra-individual monitoring of serially acquired longitudinal EEG
data.

The current method readily distinguished healthy from MS cases using
clinical EEG data in spite of obstacles faced with : \emph{i)} early
stage of the software implementation; \emph{ii)} swap in the original
purpose of the approach from monitoring serial data to cross-sectional
comparing of subjects; \emph{iii)} low information density in common
clinical EEG recordings.  We explain the analysis outcome with flow
derangement between electrical mental processes in MS secondary to
altered physical brain properties
\citep{Kisler2020,Paolicelli2021,Salim2021,Watorek2024}.  This was
newly confirmed in recent studies about disturbances of conductance
velocity \citep{Sorrentino2022,Wagner2025} although we currently offer
no method to quantify processing speed in brain states.  However, we
already know that certain feature combinations of high relevance carry
slow EEG components associated with both eye-open and eyes-closed
conditions, that appear to prominently drive, but still as part of the
entire train of feature combinations, the separation of controls from
patients (see figure {\textsf{\textbf{\ref{F2}}}}).

We speculate that even just small conductance alterations in MS will
perceptibly skew the 'electric symphony' among the many brain
activities, 'handshakes', synchronicity and interplay between
resonating loops \citep{Buzsaki2005c,Watorek2024}, all the while the
patient still \emph{compensates} without visible clinical deficit.
However, our concept about internal clockworks being disturbed between
brain activities will remain hypothetical without having access to
high-quality EEG data \citep{Formica2025,Watorek2024}.

\subsection{Significance} 

That this method succeeds on clinical-grade EEG data rests on several
facts, in the processing of \emph{complete} brain states, in the
intrinsic \emph{noise suppression} capacity, and in the characterizing
of \emph{entire} EEG recordings in form of comprehensive brain states
sequences \emph{without cleansing} noisy epochs.  The need for
artifact removal is an ongoing debate in the field while retaining the
data in the most original state is a stringent reality
\citep{Delorme2023}.  In fact, the removal of artifactual,
paradigm-related, non-brain-derived EEG components can detrimentally
affect analysis of brain physiological data \citep{McDermott2021},
just as noise filtering may lead to undesired outcomes with poorly
characterized noise patterns that are loosely related to behavior
\citep{Meisler2019}.  Experimental paradigms, however, come
necessarily with random patterns of incidental and intentional
muscular activities which surface as subject specific chance noise in
space and time.  The data processing pipeline during the brain state
detection and the later analysis phases intrinsically suppress noise
components \citep{Janoos2011,Morocz2012} and fundamentally render the
approach apt to handle clinical-grade real-life data
\citep{Janoos2013}.  This `design-specific noise robustness' lies in
the fact that brain states are generated for the entire data set and
hence hold \emph{inter}-individual, universally present features but
suppress \emph{intra}-individual periodic or permanent signal
properties.  This way do patient-specific EEG artifacts due to motor,
sensory and other effects not survive in brain states nor lead to
subgroups in the analysis nor bias the subsequently computed disease
predictor space.

Together these points highlight potential advantages of the method in
that older less-advanced EEG databases could be sensibly re-analyzed
with modern brain states analysis techniques, while collecting new EEG
data in MS or other conditions could for budgetary reasons happen on
affordable low-resolution EEG recording devices.

Having said that, only high sensor density
\citep{Formica2025,Watorek2024}, high scan frequency and rich
cognitive stimulative task conditions will leverage dense enough
information in the EEG and hence contribute to deep brain states
abstraction models.  While clinical EEG data come in their own right
only the highest-quality EEG data can deliver the fodder to study the
internal syntax inside multi-hierarchical cognetom cascade models.

\subsection{Limitations} 

First, the low spatial and temporal resolution of the analyzed EEG
data, albeit acquired at clinical grade, is an important limiting
factor whereas instead an increase in those acquisition parameters is
indeed requisite for arriving at the maximal potential of the
presented analysis technique.

Second, the two non-engaging experimental conditions, eyes-closed and
eyes-open, further limit what a more engaging cognitive session could
offer.  We expect that exposing participants to standard audio-visual
stimuli or to disease-specific experimental tasks will greatly improve
sensitivity of the approach to subject- or disease-specific patterns
in the EEG signal.

Third, the difference of mean age between the two cohorts amounts to
nine years.  We are, however, for the current retrospective analysis
not aware of a correlation between age and disease severity (EDSS or
symptoms) or disease duration.  All subjects still belong to a life
stage where we expect task-less experiments to recruit similar brain
functions and areas and hence to not age-specifically bias the brain
states predictor scale.

Fourth, the brain states epoch window kept for data-technical reasons
at two seconds width proved to work surprisingly well but does no
justice to the speed at which ephemeral electrical brain activities
operate.  Hence this window size will be optimized in future
iterations of this project with a multi-STFT approach \citep{Ali2022}.

Fifth, the current implementation of the BSD method is not suited for
the topographic visualization of the 8 brain states, as it is in the
case of the microstates method that follows a clear anatomical
preponderance \citep{Lehmann1987}.  The spatial information was for
code-technical and computational performance optimization reasons
lost.  Our current method focuses on a channel-spectral approach for
analyzing EEG data of patients with MS.

Sixth, we are unaware of the medication given to study participants
nor was this information conveyed prior \citep{Jouzizadeh2021}.
Certain drug classes are known to modify physiological brain responses
where patients with MS may indeed receive periodically benzodiazepines
\citep{vanSchependom2019} for example to control muscle spasms and
sleep disturbances, though such symptoms tend to occur at later
disease stages.  This makes a benzodiazepine induced bias in the
analysis outcome of the present study less likely in view of the short
mean disease duration of 5.1 years.  Similarly does the predominant
\emph{delta} character of the leading brain states feature
combinations (see figure {\textsf{\textbf{\ref{F2}}}}) point away from
the \emph{beta} frequency presence in benzodiazepine treated patients
with MS \citep{vanSchependom2019}.

\subsection{Future objectives} 

First we see the purpose of this publication in the introduction of a
novel hypothesis-free data-driven analysis method tested on EEG and
clinical data of patients with MS.  However another reason is our
desire to raise interest in the community towards functional brain
data in \emph{longitudinal} clinical studies in MS in the form of
\emph{serial} EEG measures \citep{Kiiski2018}.

We envision further improvements in the brain states detection (BSD)
process through widening the spectrum of methods that analyze the EEG
signal.  For example the analysis through multifractality and
persistence effectively characterizes EEG recordings of patients with
MS during eyes open and eyes closed conditions \citep{Watorek2024}.
We consider multifractality and information about long-range temporal
correlations, wavelet analysis and microstates, as further procedural
candidates that enrich the brain states feature space and enhance
model performance for time series analyses in future iterations of
this project.

Last, we hope a future MS-specific EEG repository will foster novel
quantitative neurocognitive methods on disease progression and drug
efficacy \citep{Tur2023,Montobbio2025}, similar to how the
International Progressive MS Alliance (IPMSA) \citep{Thompson2022}
catalyzes research between academia and pharmaceutical industry for
MRI data in MS \citep{Eshaghi2021,Falet2022}.

\section{Conclusion}

We introduced a novel hypothesis-free quantitative full-brain analysis
approach and successfully separated EEG data paired with clinical
evidence of patients suffering from MS from those of healthy subjects.
Thinking forward, we envision quantitative cognitive imaging using
\emph{serial} EEG data to become a future main player for
\emph{longitudinal} clinical studies where \emph{functional}
neurobiological outcome measures are key, be that for monitoring MS or
other neurological and psychiatric illnesses with prolonged time
courses.  However, as the knowledge base with cognitive imaging
techniques grows, we believe that EEG will also play a role in the
\emph{diagnosis}, \emph{subtyping} and \emph{staging} of primary and
secondary brain diseases.

\section{Declaration of conﬂicting interests}

The authors declared the following potential conflicts of interest
with respect to the research, authorship and publication of this
article: István Mórocz and Andrei Mogoutov co-founded Noisis Inc., an
early-stage startup.

\section{Funding}

The authors did not receive any financial support nor any specific
grant from funding agencies in the public, commercial, or
not-for-profit sectors for the research, authorship, and publication
of this article.

\section{Data availability statement}

Data sharing may be applicable upon request to the data controller R.K.

\newpage

\bibliographystyle{unsrtnat} 




\appendix 

\pagestyle{empty}
\section{Table of study participants (control subjects)}
\label{ST2}
\captionsetup[table]{skip=20pt}
\begin{table}[H]
  \centering
  \footnotesize\sf
  \begin{tabular}{c|ccc}
    \toprule
    index & cohort & age & sex \\
    \midrule
    0  & CT & 21 & m \\
    1  & CT & 21 & m \\
    2  & CT & 19 & m \\
    3  & CT & 29 & f \\
    4  & CT & 19 & f \\
    5  & CT & 20 & f \\
    6  & CT & 18 & f \\
    7  & CT & 19 & m \\
    8  & CT & 18 & m \\
    9  & CT & 21 & m \\
    10 & CT & 19 & m \\
    11 & CT & 19 & f \\
    12 & CT & 21 & f \\
    13 & CT & 20 & f \\
    14 & CT & 25 & f \\
    15 & CT & 18 & m \\
    16 & CT & 18 & m \\
    17 & CT & 21 & m \\
    18 & CT & 21 & m \\
    19 & CT & 18 & m \\
    20 & CT & 19 & m \\
    21 & CT & 18 & f \\
    22 & CT & 28 & m \\
    23 & CT & 19 & f \\
    24 & CT & 18 & f \\
    25 & CT & 18 & f \\
    26 & CT & 20 & f \\
    27 & CT & 18 & f \\
    28 & CT & 20 & f \\
    29 & CT & 18 & f \\
    30 & CT & 18 & m \\
    31 & CT & 21 & f \\
    32 & CT & 19 & f \\
    33 & CT & 19 & f \\
    34 & CT & 19 & m \\
    35 & CT & 20 & m \\
    36 & CT & 19 & m \\
    37 & CT & 21 & m \\
    38 & CT & 18 & m \\
    39 & CT & 20 & m \\
    40 & CT & 27 & m \\
    41 & CT & 23 & m \\
    42 & CT & 25 & f \\
    43 & CT & 19 & f \\
    44 & CT & 19 & m \\
    45 & CT & 21 & m \\
    46 & CT & 31 & m \\
    47 & CT & 28 & f \\
    48 & CT & 25 & f \\
    49 & CT & 19 & m \\
    50 & CT & 23 & f \\
    51 & CT & 19 & f \\
    \bottomrule
  \end{tabular}
  \caption{Table is continued with patient data on the \emph{next} page.}
\end{table}
\newpage

\pagestyle{empty}
\section{Table of study participants (patients with MS)}
\label{ST3}
\captionsetup[table]{skip=20pt}
\begin{table}[H]
  \centering
  \footnotesize\sf
  \begin{tabular}{c|ccc|ccp{30mm}|c}
    \toprule
    index & cohort & age & sex & \pbox{16mm}{disease\\ duration} & EDSS & symptoms & code \\
    \midrule
    
    0  & MS & 32 & f &  3 & 0   & bs,vision,sensory     &   \\
    1  & MS & 28 & f & 11 & 7   & sensory               &   \\
    2  & MS & 24 & f &  0 & 0   & bs,balance,motor      & b \\
    3  & MS & 37 & f &  2 & 1   & sensory               &   \\
    4  & MS & 40 & f & 14 & 0   & motor,sensory         &   \\
    5  & MS & 21 & f &  2 & 1   & sensory               & g \\
    6  & MS & 23 & m &  1 & 0   & vision                &   \\
    7  & MS & 31 & m &  6 & 0   & bs,sensory,motor      &   \\
    8  & MS & 29 & m &  5 & 1.5 & bs                    &   \\
    9  & MS & 27 & m &  4 & 2.5 & motor                 &   \\
    10 & MS & 27 & m &    & 0   &                       & f \\
    11 & MS & 32 & f &  1 & 0   & sensory,motor         & e \\
    12 & MS & 26 & f &  0 & 7   & motor                 & a \\
    13 & MS & 33 & m &    & 0   &                       &   \\
    14 & MS & 31 & f &    & 0   &                       &   \\
    15 & MS & 26 & f &  6 & 2   & sensory,motor         &   \\
    16 & MS & 30 & f &    & 0   &                       & d \\
    17 & MS & 20 & f &  1 & 4   & motor                 & c \\
    18 & MS & 24 & m &  0 & 1   & vision                &   \\
    19 & MS & 21 & m &    & 0   &                       &   \\
    20 & MS & 31 & m &  0 & 2   & sensory,motor         &   \\
    21 & MS & 37 & m & 15 & 4   & vision                &   \\
    22 & MS & 33 & m &  5 & 4.5 & vision,motor          &   \\
    23 & MS & 21 & m &  1 & 0   & vision,balance        &   \\
    24 & MS & 53 & m & 10 & 6   & sensory,motor         &   \\
    25 & MS & 35 & f &  8 & 5.5 & sensory,motor         &   \\
    26 & MS & 40 & f &  3 & 2   & vision,sensory,motor  &   \\
    27 & MS & 32 & f &  5 & 6   & motor                 &   \\
    28 & MS & 28 & f &  5 & 0   & vision,motor,sensory  &   \\
    29 & MS & 32 & f &  3 & 0   & vision,balance        &   \\
    30 & MS & 30 & f &  2 & 1   & sensory               &   \\
    31 & MS & 27 & f & 13 & 4.5 & vision,sensory        &   \\
    32 & MS & 26 & f &    & 0   &                       &   \\
    33 & MS & 23 & f &  1 & 0   & sensory               &   \\
    34 & MS & 32 & f &  6 & 0   & sensory,motor         &   \\
    35 & MS & 25 & f &  2 & 3.5 & vision,motor          &   \\
    \bottomrule
  \end{tabular}
  \caption{List of study participants with multiple sclerosis (MS).
    The `symptoms' column reports neurological deficits for those
    patients where such clinical information was available.  The
    related notes in the source data were of sparse and moderate
    quality.  For simplicity, the neurological deficits were binned
    into 5 categories such as [\,balance, bs, motor, sensory,
      vision\,] (with `bs' for brain stem).  The `code' column with
    alphabetic characters (a-g) stands for the seven patients that
    were found as outliers on the left side of the disease predictor
    axis marked in green letters (a-g) in figure
    {\textsf{\textbf{\ref{F3}}}}.}
\end{table}
\newpage

\section{Mean spectra for EEG frequency bands}
\label{SF4}
\begin{figure}[!ht]
  \setlength{\fboxsep}{0pt}%
  \setlength{\fboxrule}{0pt}%
  \centering
  \includegraphics[scale=.85]{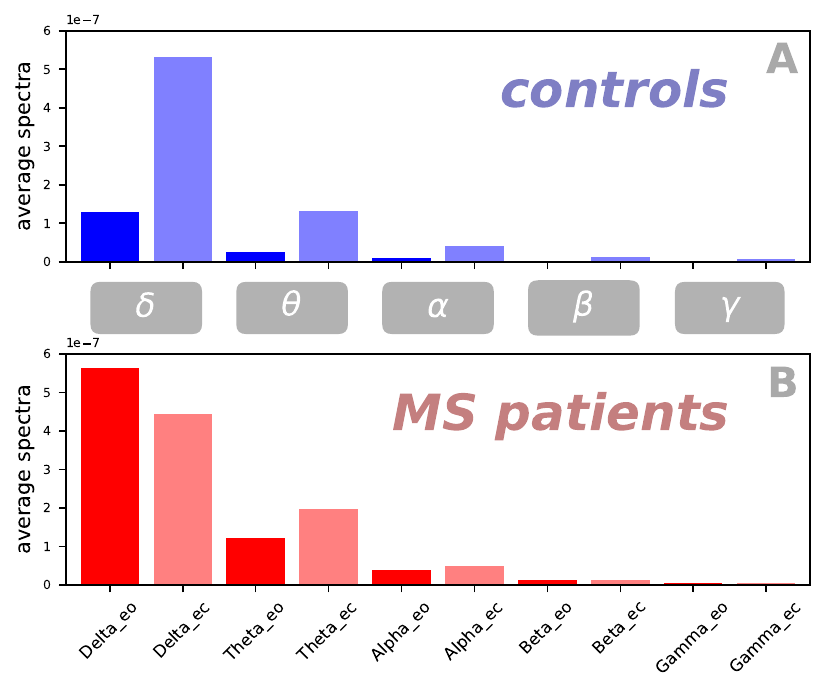}
  \caption{Graph showing mean spectra for EEG frequency bands.  The
    average for each of the five common EEG frequency bands ($\delta$,
    $\theta$, $\alpha$, $\beta$ and $\gamma$) was computed across all
    subjects, separated for control and MS cohorts, and for the two
    experimental conditions.  The eyes-open (EO) condition shows bars
    at full opacity, and the eyes-closed (EC) condition at half
    opacity.}
\end{figure}
\thispagestyle{empty}

 \newcommand{\leadingzero}[1]{\ifnum #1<10 0\the#1\else\the#1\fi}
\newcommand{\myTime}{\DTMcurrenttime} 
\newcommand{\myToday}{\the\year-\leadingzero{\month}-\leadingzero{\day}}
\newcommand\myVersion{manuscript v\,0.36}
\vfill\hfill[ \myVersion\ - \myToday\ - \myTime\ ]

\end{document}